\documentstyle[twoside,natbib,epsf]{article}

\input{ibvs2.sty}

\def\cite#1{\citealt{#1}}
\def\ibvs{Inf. Bull. Var. Stars}

\def\apj{ApJ}

\def\apss{Ap\&SS}
\def\aap{A\&A}

\def\pasp{PASP}
\def\pasj{PASJ}
\def\PublisherUAP{Tokyo: Universal Academy Press}

\begin{document}

\IBVShead{xxxx}{xx January 2002}

\IBVStitle{V893 Sco Is Not an ER UMa-Type Star}

\IBVSauth{Kato,~Taichi$^1$, Matsumoto,~Katsura$^{2,1}$, Uemura,~Makoto$^1$,}
\vskip 5mm

\IBVSinst{Dept. of Astronomy, Kyoto University, Kyoto 606-8502, Japan, \\
          e-mail: (tkato,uemura)@kusastro.kyoto-u.ac.jp}

\IBVSinst{Graduate School of Natural Science and Technology,
          Okayama University, Okayama 700-8530, Japan, \\
          e-mail: katsura@cc.okayama-u.ac.jp}

\IBVSobj{V893 Sco}
\IBVStyp{UG+E}
\IBVSkey{dwarf nova, classification}

\begintext

   V893 Sco is a recently rediscovered bright dwarf nova \citep{kat98v893sco}.
The star was subsequently found to be an eclipsing dwarf nova below the
period gap (\cite{tho99v893sco}; \cite{mat00v893sco}; \cite{bru00v893sco}).
Most recently, \citet{mas01v893sco} proposed an idea that V893 Sco
is an ER UMa-type dwarf nova.  From their analysis of the evolutionary
state of V893 Sco, \citet{mas01v893sco} proposed that all ER UMa stars
may be newly formed cataclysmic variables (CVs).

   ER UMa stars are a class of SU UMa-type dwarf novae (for a recent
review of SU UMa-type dwarf novae, see \cite{war95suuma}).  ER UMa stars
show extremely short (19--50 d) supercycles. i.e. intervals between
successive superoutbursts (e.g. \cite{kat95eruma}; \cite{rob95eruma};
\cite{nog95rzlmi}; \cite{kat96diuma}).

   \citet{mas01v893sco} analyzed the light curve from VSNET
(http://www.kusastro.kyoto-u.ac.jp/vsnet), and identified outburst
intervals of $\sim$30 d as being a supercycle.  They also identified that
{\it normal outbursts} with amplitudes of $<$1 mag every few days.
Here we report an argument against this interpretation.

   Firstly, a supercycle of ER UMa-type dwarf novae is largely occupied
by a long-lasting superoutburst \citep{kat99erumareview}.  A superoutburst
comprises 30--45\% of a supercycle (cf. \cite{rob95eruma};
\cite{kat01v1159ori}).  The outbursts occurring every $\sim$30 d last
only less than a few days (see Fig. 1), amount to a duty cycle of only
$\sim$0.1.  Furthermore, no superhumps,
which are always seen during ER UMa-type superoutbursts, have yet been
observed during these outbursts (S. Kiyota, private communication;
the observations on 1999 May 12 and 13 by \citet{mat00v893sco} were
also done during one of these outbursts, and no signature of superhumps
was observed).  These outbursts bear all characteristics of normal
outbursts rather than those of superoutbursts.

\IBVSfig{12cm}{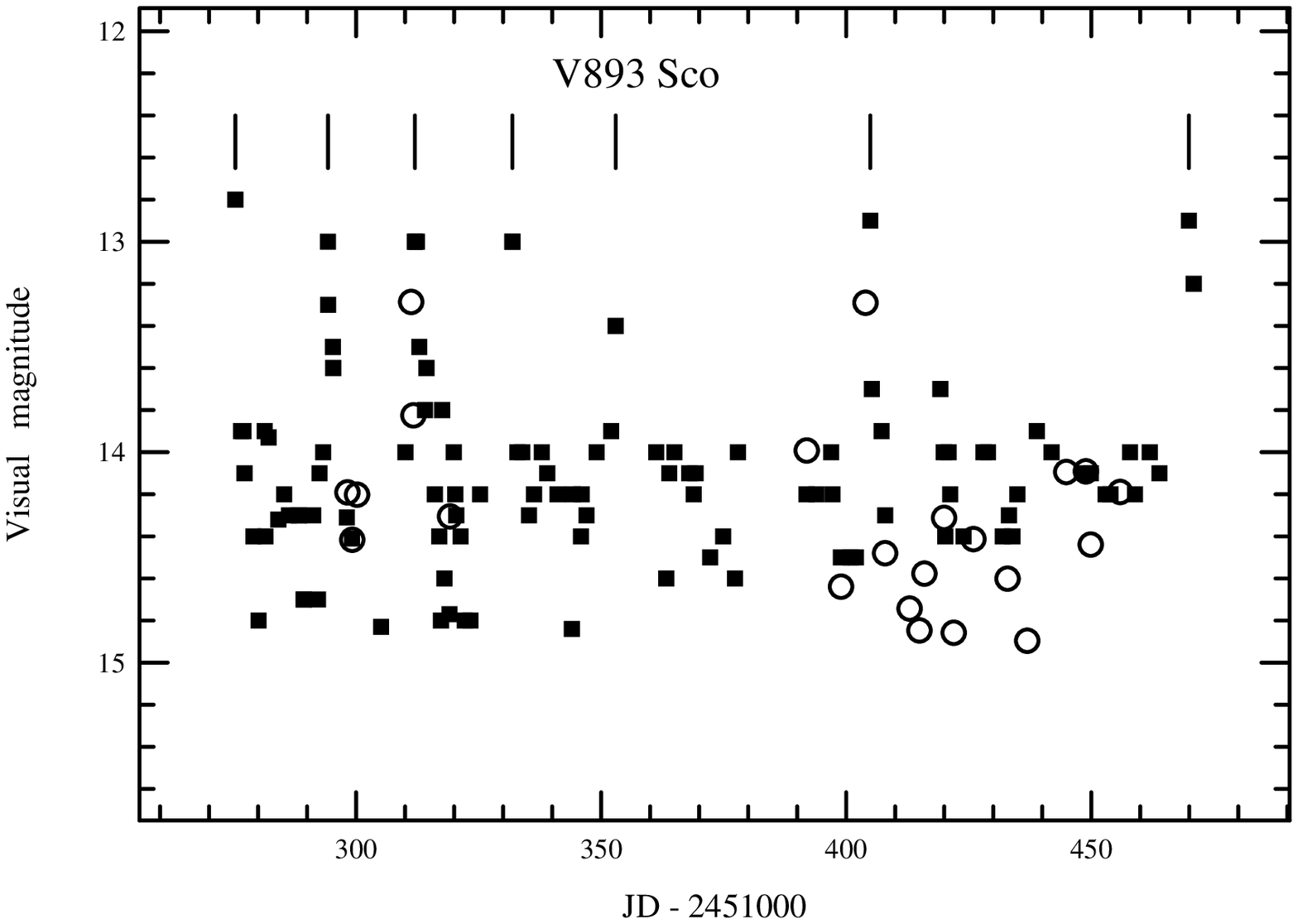}{A part of light curve of V893 Sco.
        Filled squares and open circles represent visual observations by
        VSNET observers and CCD observations by the authors.
        CCD observations were carried out using a 25-cm Schmidt-Cassegrain
        telescope and an ST-7 CCD.  The zero-point adjustment was made
        using Kiyota's CCD observation (private communication).
        The typical errors of observations are 0.2 mag (visual), 0.1 mag
        (CCD).
        Short outbursts (normal outbursts) are marked with vertical bars.
        The light curve is unlike those of ER UMa stars in that V893 Sco
        lacks long-lasting superoutbursts.
}

   Secondly, what were referred to as {\it outbursts with amplitudes of $<$1
mag every few days} in \citet{mas01v893sco} is not evident as shown in
Figure 2.  Both visual and CCD observations show only {\it irregular}
variations which are frequently met in CVs.  Period analysis yielded
not significant periodicity.  The presence of such strong quiescent
variations in V893 Sco was also independently confirmed by
\citet{bru00v893sco}.  These variations can to be better understood
as an enhanced activity sometimes observed in high-inclination systems
\citep{kat01v2051ophiyuma}.

\IBVSfig{10cm}{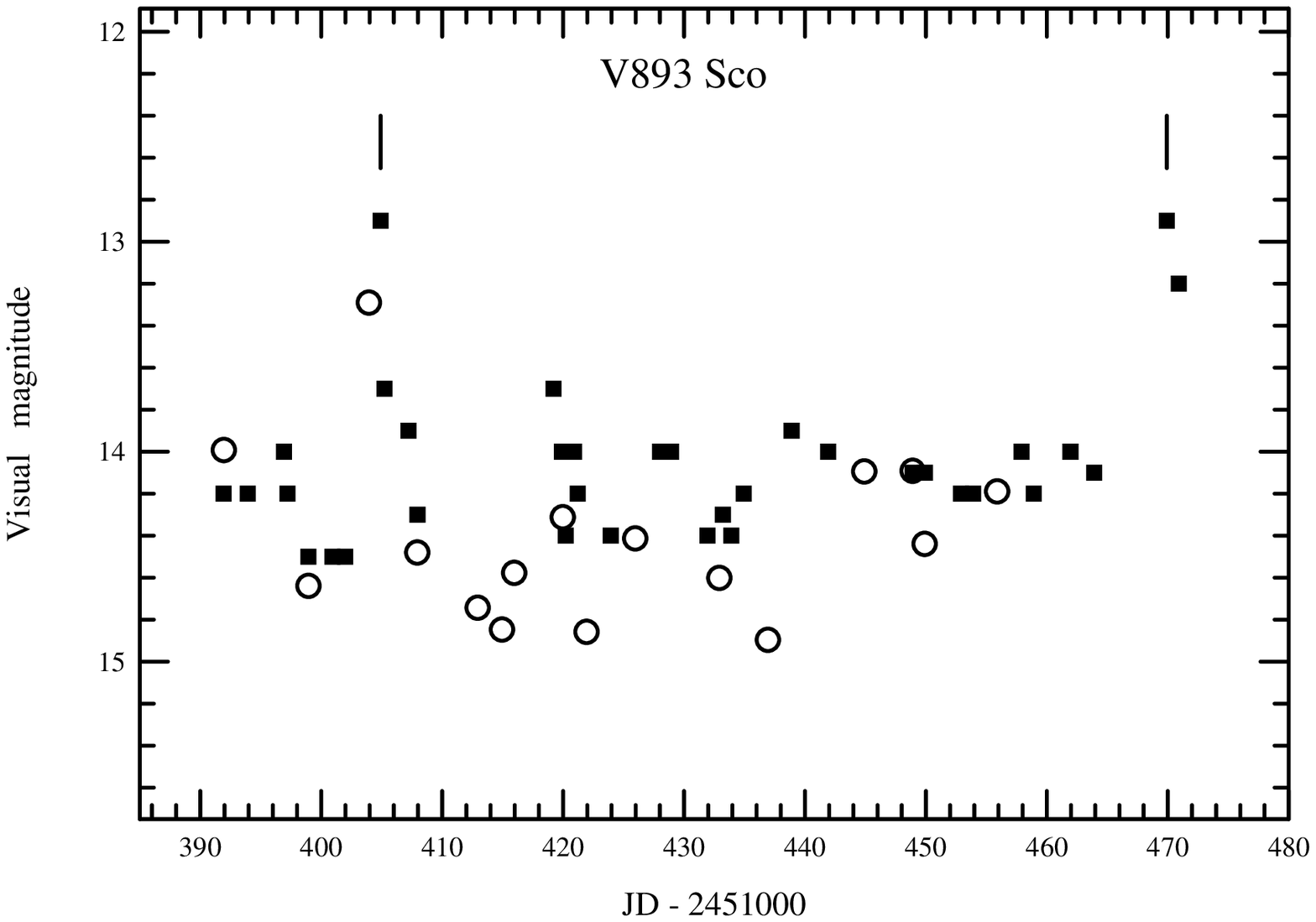}{Enlargement or Fig. 1.  The symbols are the
        same as in Fig. 1.  Only irregular variations were observed
        between outbursts.
}

   From these findings, we conclude that V893 Sco bears no similar
characters with ER UMa stars, and that the arguments in \citet{mas01v893sco}
need to be reconsidered.

\vskip 3mm

We are grateful to Rod Stubbings, Berto Monard, and Andrew Pearce,
who reported vital observations to VSNET.
Part of this work is supported by a Research Fellowship of the
Japan Society for the Promotion of Science for Young Scientists (KM, MU).
This work is partly supported by a grant-in aid (13640239) from the
Japanese Ministry of Education, Culture, Sports, Science and Technology.

\end{document}